\documentclass[aps,prl,twocolumn,showpacs,amsmath,amssymb,superscriptaddress]{revtex4}

\usepackage{graphicx}
\usepackage{bm}
\usepackage{color}
\usepackage{ulem}
\usepackage{textcomp}

\begin{document}

\newcommand{\be}{\begin{eqnarray}}
\newcommand{\ee}{\end{eqnarray}}


%
%
%
%
%


{\bf Planet, Santucci, and Ort\'{\i}n Reply:}

 

In Ref.\ \cite{Planet-et-al-09} we reported that both the size and duration of the global avalanches observed during a forced imbibition process follow power law distributions with cut-offs,  
$P_x(x) = a_x x^{-m_x} {\cal G}_x(x/x_c)$.   
While the exponent of the power law appears robust within the quoted error bars, the 
cut-off of the pdf's depends on experimental control parameters
such as the injection rates $v$. 


When adimensionalising the variables,  $u = x/\langle x \rangle$, we observed a collapse of the various pdf's $P_u(u)$. Thus, the exponent $m_x$ should be found equal to one, as explained in \cite{Pruessner-10}.
Indeed, we observed clearly an average power law exponent $\alpha = 1.00 \pm 0.06$ for the avalanche size pdf's. However, for the avalanche duration we reported a slightly larger value within the large dispersion. This can be attributed to the poorer statistics for the avalanche duration, as  we explained in \ \cite{Planet-et-al-09}, 
affecting the quality of the collapse and/or the fitting procedure.  This illustrates the difficulty of extracting accurate values of the power law exponents with experimental data with a limited statistics.    
%
\begin{figure}
   \includegraphics[width=0.5\textwidth]{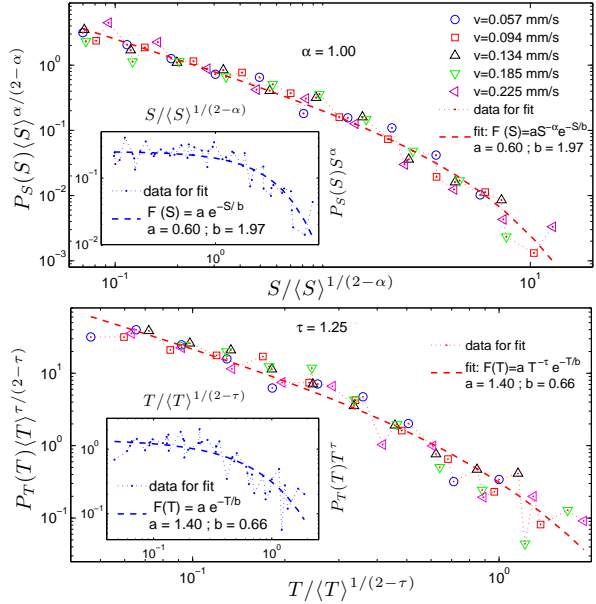}
   \caption{(Color online) Top: Statistical distributions of avalanche sizes $S$ for various injection rates (clip level C=0). 
   The main plot shows $P_S(S) \langle S \rangle^{\alpha/(2-\alpha)}$ vs $S/ \langle S \rangle^{1/(2-\alpha)}$, and the inset   $P_S(S) S^{\alpha}$ vs $S/ \langle S \rangle^{1/(2-\alpha)}$, both computed for the value $\alpha=1.00$ that provides the best collapses. Bottom: Corresponding collapses of the pdf's of the avalanche durations $T$, for the value $\tau = 1.25$.}
\end{figure}

In order to extract reliable exponents, the right procedure --as described in \cite{Pruessner-10}-- is
to find the power law exponent $m^*$ that provides the best collapse of  $Y = P_x(x) \langle x \rangle^{m/(2-m)}$ as a function of $X = x/\langle x \rangle^{1/(2-m)}$.    
In order to quantify the quality of this collapse, first, we have fitted $Y_{fit} = AX^{-m} \exp(-X/B)$ to the set of our experimental data, varying systematically the exponent $m$ and leaving A and B as free parameters. Then, we have computed the error $\epsilon = \sum_i[\log(Y_i) - \log(Y_{fit})]^2$ as a function of $m$. 
For various velocities (clip level C = 0), for the size distribution,  the minimum value of $\epsilon$ (thus, the best collapse) is obtained for $\alpha = 1.00 \pm 0.15 $,  
while for the avalanche durations we obtain  $\tau = 1.25 \pm 0.25$, as shown in Fig. 1.   
The insets display the collapse $P_x(x) x^m$ as a function of $x/\langle x \rangle^{1/(2-m)}$,
showing that the scaling function is very well approximated by a decaying exponential.
%
Then it is not difficult to show that the joint distribution of sizes and durations can be properly analyzed using $u' = S/\langle S \rangle^{1/(2-\alpha)}$ and $w' = T/\langle T \rangle^{1/(2-\tau)}$, with the values of the exponents previously found. We show in Fig. 2 that $u' \propto w'^{\gamma}$, where $\gamma = 1.33 \pm 0.12$. 
\begin{figure}
 \includegraphics[width=0.5\textwidth]{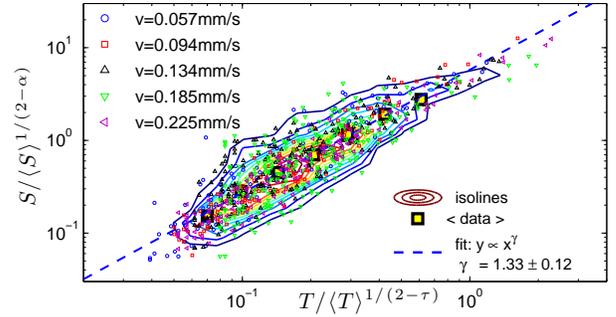}
 \caption{(Color online) Isolines of the joint distributions $u'$ vs $w'$, with a fit to the crest giving the value $\gamma = 1.33 \pm 0.12$.}
\end{figure}

It is important to notice that the values obtained here are in agreement with the original ones obtained by a direct fitting of the pdf's of the raw and dimensionless data. Since the actual exponents $m$ are close to one  $P(x/\langle x\rangle) = P(x) \langle x\rangle  \simeq  P(x) \langle x \rangle^{\frac{m}{2-m}}$ and  $x/\langle x \rangle \simeq x/\langle x \rangle^{\frac{1}{2-m}}$ , and due to both the experimental noise and limited statistics, the collapses previously observed  were reasonably good and nearly indistinguishable from the present ones.

Finally, Ref.\ \cite{Pruessner-10} suggests that the problem of forced--flow imbibition might belong to the quenched Edwards--Wilkinson (QEW) or equivalently the C-DP universality class \cite{Alava-02}, a conjecture based on a possible similarity of the values of the exponents. This should be taken with some caution, however, because forced--flow imbibition is a {\sl non-local} process \cite {Alava-04}, while the QEW interfacial equation describes a {\sl local} interfacial dynamics. Moreover, the various values of the exponents reported here are in very good agreement with the ones obtained from phase--field simulations of a {\sl non-local} interfacial process \cite{Pradas-PhD-09}.



\begin{references}
\bibitem{Planet-et-al-09}
R. Planet {\it et al.},
Phys. Rev. Lett. {\bf 102}, 094502 (2009).

\bibitem{Pruessner-10}
G. Pruessner, Comment LPK1047 
submitted to Phys. Rev. Lett., February 2010.

\bibitem{Alava-02} 
M. Alava and M.A. Mu\~noz , Phys. Rev. E, {\bf 65}, 026145, (2002).

\bibitem{Alava-04} 
M. Alava et al, Adv. Phys. {\bf 53}, (2), 83-175, (2004).

\bibitem{Pradas-PhD-09}
M. Pradas, {\it Interfaces in disordered media.
Scaling growth, avalanche dynamics, and microfluidic fronts}.
PhD Thesis, University of Barcelona (2009).







\end{references}
\end{document}